\pdfoutput=1
\documentclass[conference]{IEEEtran}
\IEEEoverridecommandlockouts
\usepackage[moderate,tracking=normal]{savetrees}
\usepackage[backend=bibtex,giveninits=true,hyperref=true,maxnames=1,minnames=1,natbib=true,style=savetrees,sorting=none]{biblatex}
\usepackage{amsmath,amssymb,amsfonts}
\usepackage{algorithmic}
\usepackage{graphicx}
\usepackage{textcomp}
\usepackage{xcolor}
\usepackage{tabularx}
\usepackage{booktabs}
\usepackage{multicol}   
\usepackage{float}
\usepackage{csquotes}
\usepackage{hyperref}

\definecolor{ForestGreen}{RGB}{34,139,34}\usepackage[ruled,vlined]{algorithm2e}
\usepackage{makecell}
\def\BibTeX{{\rm B\kern-.05em{\sc i\kern-.025em b}\kern-.08em
    T\kern-.1667em\lower.7ex\hbox{E}\kern-.125emX}}
\usepackage[english]{babel}

\usepackage{mathtools}
\usepackage{lipsum}
\usepackage{paralist}
\usepackage{enumitem}
\usepackage{subcaption}
\usepackage{nicematrix}
\usepackage{colortbl}%
\usepackage{multirow}
\usepackage{footmisc} 
\usepackage[para,online,flushleft]{threeparttable}
\usepackage{hyperref}
\usepackage{caption}

\usepackage{xcolor}

\newcommand{\linebreakand}{%
  \end{@IEEEauthorhalign}
  \hfill\mbox{}\par
  \mbox{}\hfill\begin{@IEEEauthorhalign}
}
\makeatletter 
\let\myorg@bibitem\bibitem
\def\bibitem#1#2\par{%
  \@ifundefined{bibitem@#1}{%
    \myorg@bibitem{#1}#2\par
  }{%
    \begingroup
      \color{\csname bibitem@#1\endcsname}%
      \myorg@bibitem{#1}#2\par
    \endgroup
  }%
}
\newcommand*{\bibitem@test}{blue}    
\newcommand*{\bibitem@green}{green}  
\makeatother 
\makeatletter
\newcommand\footnoteref[1]{\protected@xdef\@thefnmark{\ref{#1}}\@footnotemark}
\makeatother
\makeatother
\newcounter{myenumi}
\newenvironment{myenumerate}
{
    \begin{enumerate} 
    \setcounter{enumi}{\value{myenumi}}
}
{
    \setcounter{myenumi}{\value{enumi}}
    \end{enumerate}
}
\begin{document}
\title{Learning Cellular Coverage from Real Network Configurations using GNNs\\}


\author{
    \IEEEauthorblockN{Yifei Jin}
    \IEEEauthorblockA{
     \textit{KTH \& Ericsson Research} \\
        Stockholm, Sweden \\
        \url{yifeij@kth.se}
    }

\and

\IEEEauthorblockN{Marios Daoutis}
\IEEEauthorblockA{\textit{Ericsson Research} \\
Stockholm, Sweden \\
\url{marios.daoutis@ericsson.com}} \\

\and
    \IEEEauthorblockN{Šarūnas Girdzijauskas}
    \IEEEauthorblockA{
        \textit{KTH} \\
        Stockholm, Sweden \\
        \url{sarunasg@kth.se}
        }

\and

\IEEEauthorblockN{Aristides Gionis}
    \IEEEauthorblockA{
        \textit{KTH} \\
        Stockholm, Sweden \\
        \url{argioni@kth.se}
        }
}

\maketitle
\begin{abstract} 
Cellular coverage quality estimation has been a critical task for self-organized networks. In real-world scenarios, deep-learning-powered coverage quality estimation methods cannot scale up to large areas due to little ground truth can be provided during network design \& optimization. In addition, they fall short in producing expressive embeddings to adequately capture the variations of the cells' configurations. To deal with this challenge, we formulate the task in a graph representation and so that we can apply state-of-the-art graph neural networks, that show exemplary performance. We propose a novel training framework that can both produce quality cell configuration embeddings for estimating multiple KPIs, while we show it is capable of generalising to large (area-wide) scenarios given very few labeled cells. We show that our framework yields comparable accuracy with models that have been trained using massively labeled samples. 
\end{abstract}

\begin{IEEEkeywords}
Self-supervised Learning, Graph Neural Network, Cellular Coverage Estimation, Few-shot Learning 
\end{IEEEkeywords}

\section{Introduction}
\label{sec:introduction}
Estimating mobile users' Quality of Service (QoS) metrics in cellular networks has been a long-studied problem, in the context of network designing and optimization. 
As we approach 6G telecommunication networks, these are expected to be self-aware~\cite{chaoub2021self} (i.e., Estimating service QoS-related Key performance indicator (KPI) based on its configurations), paving the way towards the goal of Self-Organized Network (SON).
Estimating radio-coverage-related QoS still remains a challenge, even for cell-level~\cite{balevi2019online}~\footnote{We define the term 'cell' as Evolved Universal Terrestrial Radio Access (eUtran)-Cell, i.e.: the cell's coverage is denoted by one single carrier frequency's coverage.}, mainly due to the complexities that emerge during radio wave propagation. More specifically, to estimate a site with multiple cells' QoS KPI is NP-hard, whereas it is NP-complete if many sites with heterogeneous inter-cell relations exist.\\
\begin{figure}[ht]
     \begin{subfigure}[h]{0.13\textwidth}
         \includegraphics[width=\textwidth]{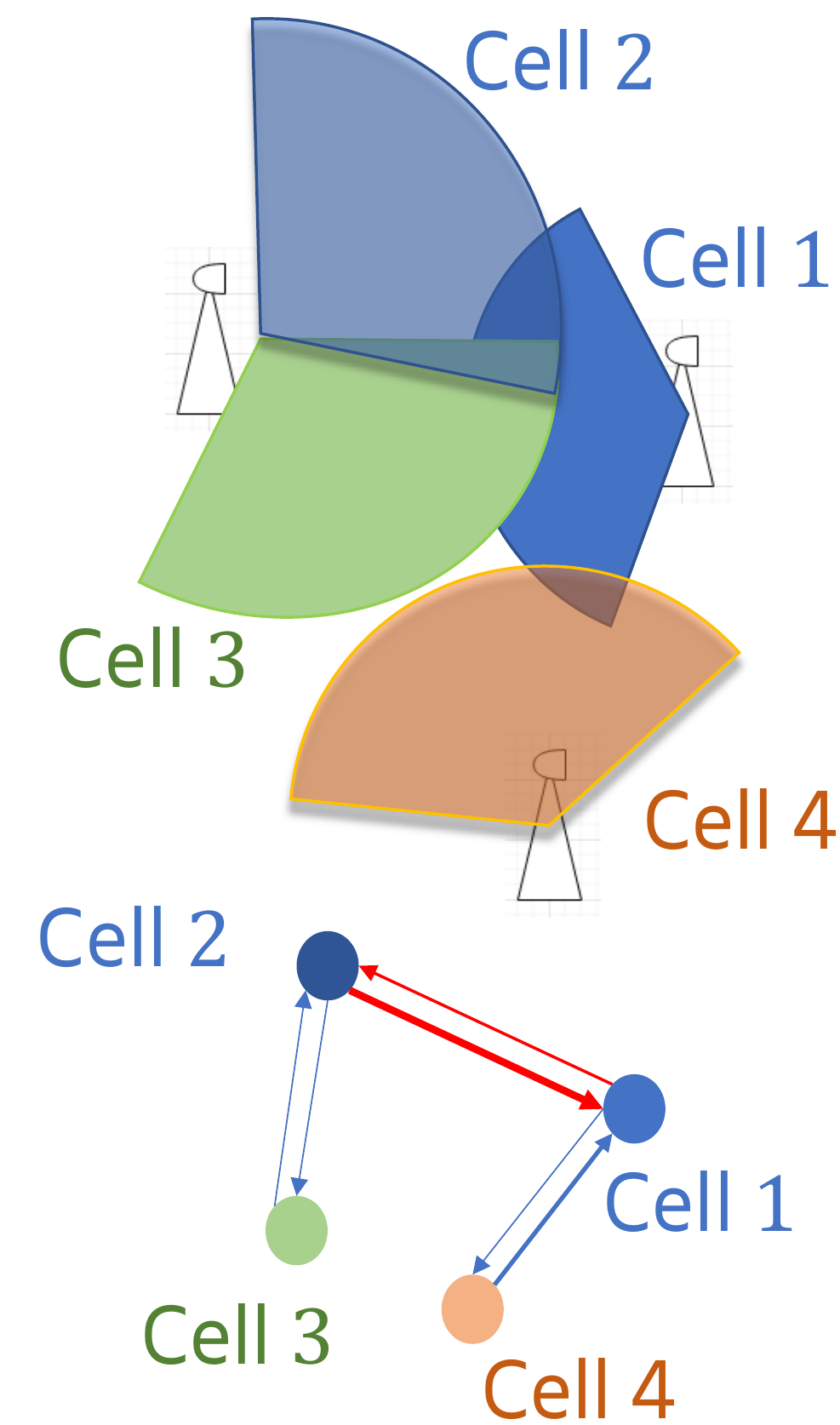}
         \caption{}
         \label{fig:cellrep}
     \end{subfigure}%
     \begin{subfigure}[H]{0.37\textwidth}
        \includegraphics[width =\textwidth]{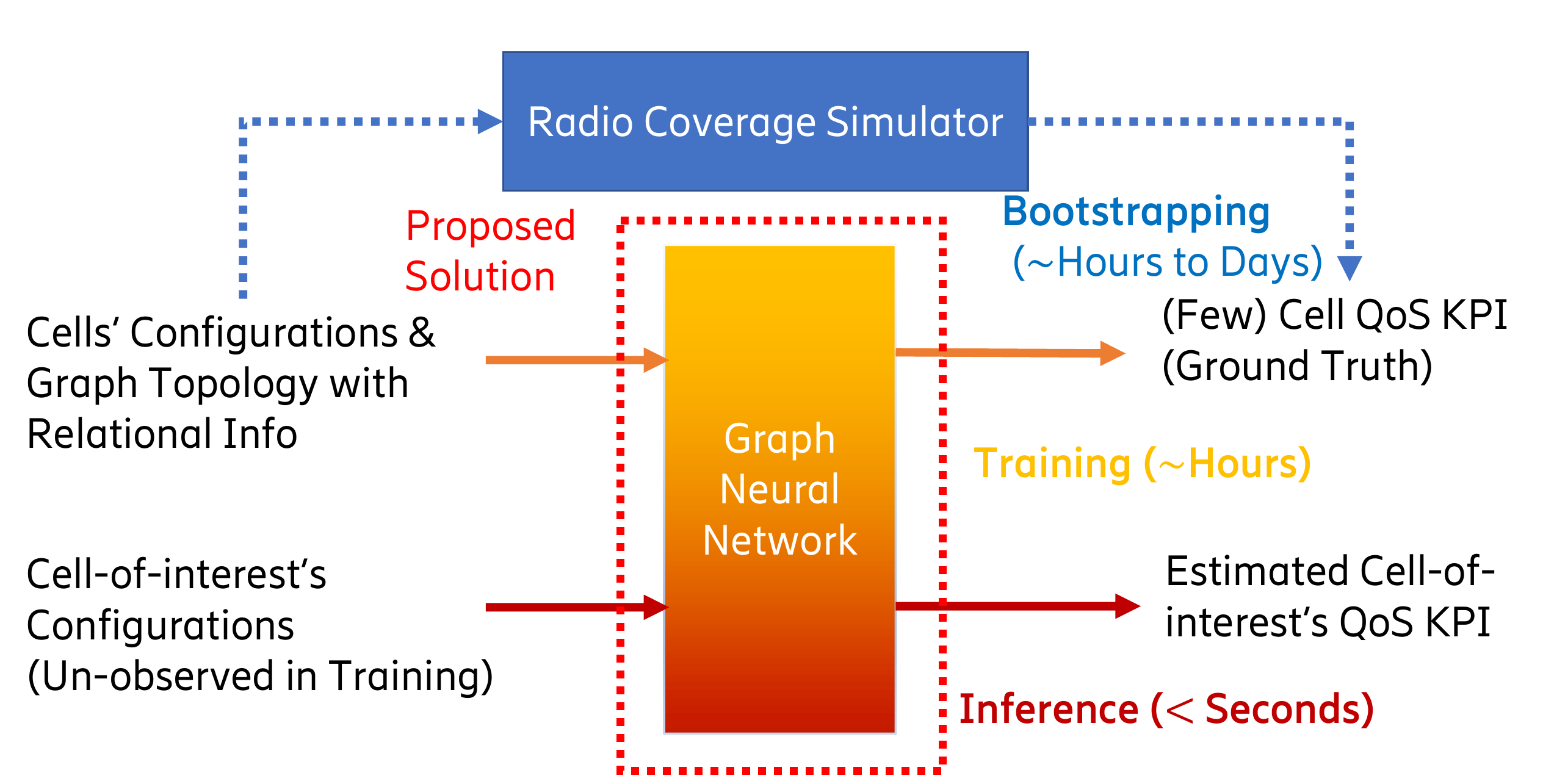}
        \caption{}
        \label{fig: solutionrep}
    \end{subfigure}
    \caption{$(a)$ Example of a graph representation of 'inter-cell' relations. Each vertex represents a cellular coverage, color-coded with its carrier frequency (Blue/Green/Orange). Each directed edge denotes an interfering (Red)/complementing (Blue) impact posed to the destination vertex, where edge strength denotes the strength of impact. $(b)$ A schematic overview of the proposed solution. Bootstrapping time ranges from hours to days, training time needs hours to complete, while test/validation time is in the range of a few seconds.}
\end{figure}
Conventional methods for such problems are often based on logic-based simulations, where the simulator integrates complex radio propagation models and computes the coverage quality on a `per-cell' basis. The simulator, then, integrates the results of each cell to produce the area's estimation.
Despite the high accuracy of such approaches, these are usually computationally very expensive and in some cases infeasible for cellular coverage computations at scale.
Due to recent advances in deep learning (DL), graph neural networks (GNN) have been proposed as an alternative way to tackle in a time- and computation-efficient way, such computationally difficult sub-problems, such as that of estimating a single cell's coverage KPI~\cite{thrane2020deep,cui2019spatial}, estimating scenarios where a cell holds multiple users~\cite{rajapaksha2021deep}, as well as estimating multiple cell-user pairs which co-exist in the same scenario~\cite{shen2020graph}. 
These holistic `inter-cell' relations (illustrated in Fig.~\ref{fig:cellrep}), though crucial during network design \& optimization, need to be learned.\\
In this paper, we study the KPI estimation problem in the context of a realistic formulation, where we aim to: 
\begin{inparaenum}[\itshape (i)\upshape]
    \item learn to estimate user QoS KPI measurement on a `per-cell' basis, where each cell's configuration is independently devised and fine-tuned by field experts, and
    \item use a minimum amount of ground truth (i.e., KPI measurement), which is costly to acquire.
    \item capture the cross-cell impact of configuration parameters through graph representation, formulated by inter-cell relation, which has not been addressed in previous works.
\end{inparaenum}
A schematic representation of the proposed solution is shown in Fig.~\ref{fig: solutionrep}.

Recent DL methods have focused on predicting QoS KPIs of single/multiple cells from the same site, using simulation, crowd-sourcing~\cite{pimpinella2021crowdsourcing}, or road test data (usually collected at a small scale)~\cite{narayanan2020lumos5g}. However, for cell-level QoS KPI prediction, network measurements are required in both quality and quantity as ground truth.
Mobile service providers expect that such a model can: \begin{inparaenum}
    \item learn from network configurations and their respective measurements in real-world deployments, and  
    \item scale-up from measurements of a few cells, instead of densely running road measurement tests or frequently triggering measurement reports for the serving users
    across many cells.
\end{inparaenum} 
On the other hand, properly formulating the QoS KPI estimation task remains an open problem in representation learning, with only few papers~\cite{jin2021graph, balevi2019online} adopting representations for modeling 'inter-cell' relations.
Even fewer papers~\cite{dreifuerst2021optimizing} have further considered heterogeneity (e.g: varying inter-site distance) in cells' representation, which is common in real-world scenarios. A successful formulation of representation would require: \begin{inparaenum}
    \item great expressivity\footnote{The ability of GNN to differentiate small perturbations on node attributes.} in terms of cell configuration parameters \& 'inter-cell' relations.
    \item the ability to perform well with a limited amount of ground truth/measurements, i.e. as for example in few-shot learning (FSL) settings.
\end{inparaenum} 
Leveraging inter-cell relation with proper representation in a realistic setup using few cells' network measurements, so far, has not been included in the problem formulation of state-of-the-art studies. 

The remaining sections are organized as follows: In Sec.~\ref{sec:related} we discuss the related publications in the domain of QoS KPI estimation of GNNs and DL in the cellular network, while in Sec.~\ref{sec:formulation} we present our problem formulation in more detail. In Sec.~\ref{sec:method} we present our methodology followed by Sec.~\ref{sec:exp}, which contains the experimental evaluation of our algorithm. Finally, Sec.~\ref{sec:conclusion} concludes with a discussion of our main contributions and considerations for future work.
\section{Related Work}
\label{sec:related}
Cell configuration parameters are usually devised and fine-tuned by field experts in real-world radio network deployments. Besides the domain knowledge involved in the configuration process, it is considered a laborious task which cannot be easily automated due to the complex relations and interactions between the networks nodes (which is neglected in many previous work~\cite{margaris2022hybrid,kaya2021deep}). Furthermore, subtle configuration changes may considerably affect the network performance through characteristics such coverage and interference (e.g., cell edge performance~\cite{you2011cell}). Here, the aim is to produce a model that has learned the associations and relations between cell configuration parameters and their QoS KPI metrics. 

Since we need to be able to capture subtle configuration changes, it is essential to employ a representation that has great expressivity. Furthermore, we need to be able to represent heterogeneous data, while learning their relations. GNN satisfies all the above requirements. Our proposed solution models the QoS KPI prediction as a \textit{node-attribute prediction task}.
Many recent advances (e.g.: \cite{xu2018powerful}) in GNNs provide the needed expressivity of the node/edge attributes, which further permits the GNN to learn heterogeneous configuration parameters. On the other hand, graph representations are getting popular in the domain of radio networks. In the context of device-to-device communication, GNNs are used to perform link-level KPI estimation~\cite{shen2020graph,zhang2021scalable,shen2022graph}. However, none aforementioned work has addressed `inter-cell' relations with respect to coverage QoS estimation, which poses a cell/system level KPI prediction.

There exist previous studies (e.g.,~\citet{wang2022spatial}) which address `inter-cell' relations through a graph representation (geo-proximity for the GNN-powered `inter-cell' handover prediction problem). However, these do not consider important aspects such as the heterogeneity of the relations, relations beyond geographically adjacent cells, and spatial orientation of antenna, all of which impact the `inter-cell' relations.
Moreover, cellular coverage embeddings have been used together with graph attention to capture the `inter-cell' relations, for the problem of antenna tilt optimization \cite{jin2021graph}. This is done in a multi-agent reinforcement learning framework, with identical action space for all antennas. However, the authors have not considered few-shot learning, while the results reported cannot be applied to real-world scenarios. In our experiment, we include network configurations crawled from real networks. 
The data-sets~\footnote{\label{note1} Detailed information regarding the dataset, training parameters and respective simulation setup will be presented in Appendix:\url{https://github.com/bluelancer/GNN4NDOSuppliment}} pose a diverse and unbalanced feature distribution, as each set of configuration parameters is fine-tuned by human experts.

One can consider auxiliary property prediction as a pretext-task in Self-Supervised Learning (SSL) scheme to overcome the data constraint. Formalized by~\citet{liu2022graph}, the auxiliary property can incorporate domain knowledge that helps GNN backbone to learn transferable insights. Given such benefit, there exist two open problems: 
\begin{inparaenum} 
    \item Formulate a proper auxiliary property.
    \item Select a suitable training scheme.
\end{inparaenum}
Some recent works (e.g.,~\citet{jin2021node}) augment auxiliary properties based on embedding's pairwise similarities. While pre-training (PT) toward such tasks can be generalized as a pre-clustering, such a scheme constrains the expressiveness of the produced embedding. 

\citet{hu2020gpt} share a training framework similar to the one we propose, which takes graph generation as pretext-tasks and produces pre-trained embeddings that can be generalized by fine-tuning  some ground truth labels. While this method is not suitable for cellular network since: \begin{inparaenum} [\itshape (i)\upshape]
    \item One can easily determine the interfering/complementing relation by each cell's carrier so that little supervision can be expected. 
    \item This method requires more than 10\% of data to be labeled in fine-tuning. This means hundreds of cell measurements, which exceeds the scale of the largest road test dataset to our best knowledge.
\end{inparaenum}
Beyond the novelty mentioned in Sec.~\ref{sec:introduction}, we demonstrate the potential for GNN and SSL to be used together in a
practical setting: through a novel formulation of auxiliary property inspired by network geometry~\cite{chen2018stochastic,cho2022simultaneous}, bringing down the labeling requirement to 2.5\%.
\section{Problem Formulation}
\label{sec:formulation}
\begin{figure*}
    \centering
    \includegraphics[width=0.8\textwidth]{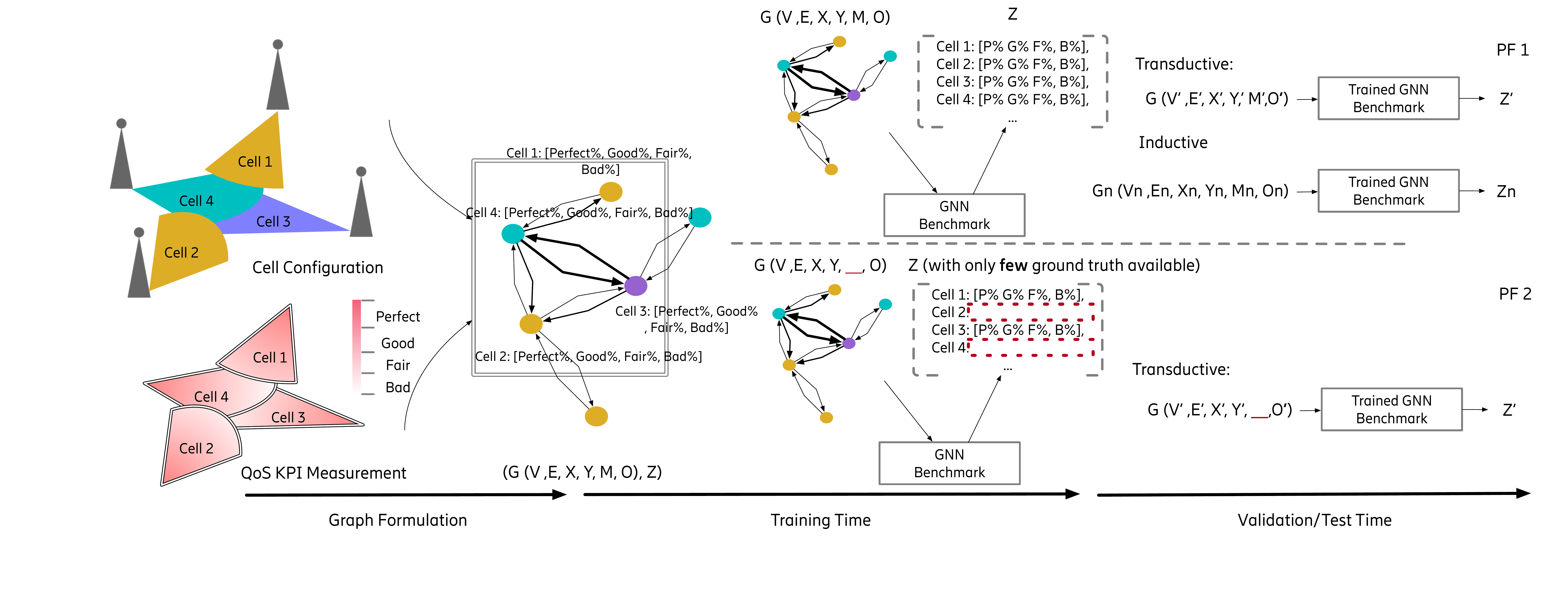}
    \vspace{-0.6cm}
    \caption{Problem formulation of coverage QoS KPI estimation. In graph formulation, we color-coded the sector (and its representing node) by its carrier frequency in a one-hot manner. Additionally, we include the cell configuration, 'inter-cell' relation, and QoS KPI measurement as nodes' and edges' attributes, and node-level ground truth, $X$, $Y$, and $Z$, respectively. \ref{form:full} trains the GNN benchmark models in training time against sufficient measurement as ground truth. \ref{form:real} only trains against very few shots of measurements as ground truth, with no measurement $M$ feature provided as node features. In test/validation time, for \ref{form:full} \& \ref{form:real} Transductive setting, the test set includes unseen cells during training time from the same dataset/area. For \ref{form:full} Inductive generalization settings, the test set include unseen scenarios with different topology, i.e. $G(V',E',X',Y',M',O')$ during training time from other dataset/area, i.e. $G_n(V_n,E_n,X_n,Y_n,M_n,O_n)$.}
    \label{fig:PF}
\end{figure*}

We consider the task of estimating the QoS KPI for cell coverage in this paper, which includes two target parameters, \textit{Signal to Interference \& Noise Ratio} (SINR) and \textit{Channel Quality Indicator} (CQI) for each cell.
\begin{enumerate}
    \item The ground truth $Z$ includes a four-dimensional array, denoting the \textit{Perfect}, \textit{Good}, \textit{Fair} and \textit{Bad} percentage of area, within the cell's coverage concerning the query KPI. 
\end{enumerate}
The input to the problem assumes an attributed directed graph representation $G(V,E,X,Y,M,O)$ that includes:
\begin{enumerate}[resume]
    \item A set of nodes $V$, where each node $v_i \in V$ denotes a cellular coverage. 
    \item A set of node attributes $X$ denoting cells' configurations, where cell $v_i$'s set of configuration is denoted as an array $x_i \in X$. For $k$-th elements in $x_i$ denotes a configuration parameter (in scalar) $x^k_i$ . 
    \item A set of directional edges $E$ that connect $V$, denoting the cell relation. Each edge $e_{ij}$ connects from source $v_i$ to destination $v_j$.
    \item A set of edge attributes $Y$, denoting the property of 'inter-cell' relations. For specific edge $e_{ij}$, the edge's property is denoted as $y_{ij} \in Y$, pose from source node $v_i$ to destination node $v_j$. Each $y_{ij}$ includes a three-dimensional array, including 
    a one-hot encoded scalar that denotes the 'inter-cell' relation type within $\{\textit{interfering}, \textit{complementing}, \textit{both}\}$,
    a numerical scalar denotes the strength of the relation, 
    and a numerical scalar that denotes geographical distance between $v_i$ and $v_j$.
    Knowing the relation is asymmetrical as $v_i$ could interfere with the whole coverage of $v_j$, while $v_j$'s interfering coverage may only take up a small portion of $v_i$'s coverage area. 
    \item A set of measurement features $M$, which are expensive to retrieve as mentioned previously. $M$ includes \textit{Received Signal Strength Indicator} (RSSI) in a four-dimensional array, denoting the \textit{Perfect}, \textit{Good}, \textit{Fair} and \textit{Bad} percentage of area. 
    $m_i \in M$ is considered as additional node attributes of $v_i \in V$.
    \item A set of augmented geometric features $O$, inspired by recent network geometry studies \cite{chen2018stochastic} to model the orientation and shape of the coverage. 
    $\mathbf{g}_{ij} \in O$ are considered as additional edge attributes of $e_{ij}$ or a novel pretext-task in a different formulation of the problem. The detailed formulations present as follows.
\end{enumerate}
We consider two problem formulations (PF) in this paper:
\begin{enumerate}[label=\textbf{PF\arabic*}]
    \item\label{form:full} we include all possible features (i.e. $G(V,E,X,Y,M,O)$, with abundant $Z$) in training time, to benchmark the upper bound of baseline model performance on estimating $Z$. 
    \item\label{form:real}We include only realistic features (i.e. $G(V,E,X,Y,\_,O)$, with few-shot $Z$) in training time, to benchmark and minimize
    the performance degradation
    on estimating $Z$ in FSL.
\end{enumerate}
We report both Transductive and Inductive generalization setting performance in \ref{form:full}, and report only the Transductive setting performance in \ref{form:real}. The motivation for a second PF is three-folded: \begin{inparaenum} \item $M$ includes RSSI which is not included in signaling from eNB to core network, making this expensive to collect in the real network. \item A huge amount of ground truth is required for \ref{form:full}. \item In Sec.~\ref{sec:exp}, we see that the GNN does not learn $O$, which motivates us to use SSL to reinforce the training.\end{inparaenum} A detailed illustration of the problem formulation is provided in Fig.~\ref{fig:PF}

\section{Methodology}
\label{sec:method}
\begin{figure}
     \begin{subfigure}{0.2\textwidth}
         \includegraphics[width=\textwidth]{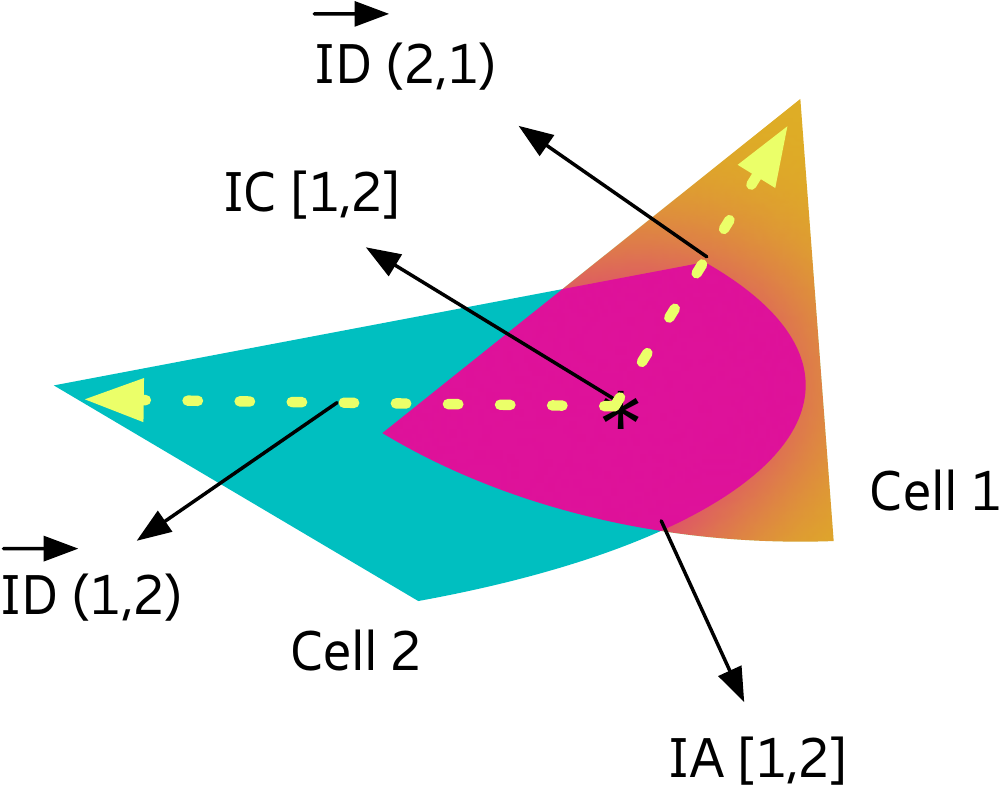}
         \caption{}
         \label{fig:geom}
     \end{subfigure}
     \begin{subfigure}{0.3\textwidth}
        \includegraphics[width =\textwidth]{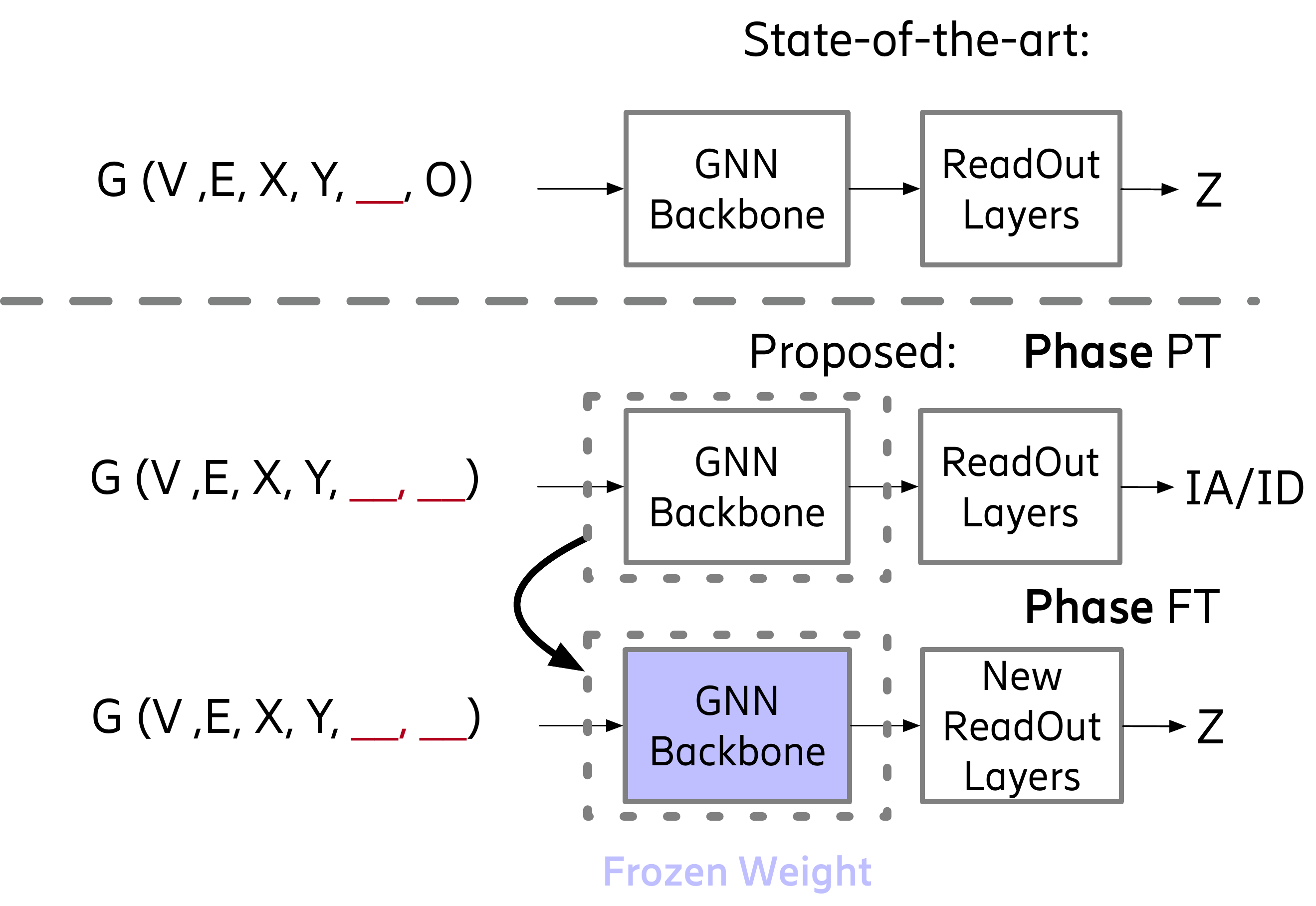}
        \caption{}
        \label{fig: proposed}
    \end{subfigure}
    \caption{($a$), Illustrative representation of formulated geometric feature $O$ ($b$), Schematic Representation of Proposed Self-supervised PT solution (i.e.: Algorithm.~\ref{alg:SSL}) in~\ref{form:real}, Compared with State-of-the-art solution.}
\end{figure}

To prove the argument on \textbf{GNN's capability of learning the inter-cell relation for QoS KPI estimation} in the above problem formulation, we benchmark many state-of-the-art models that are reported to be leading in coverage QoS KPI estimation task with different expressivity in Sec.\ref{sec:related}: 
GAQ/GAT, GINE, WCGCN and multilayer perception (MLP), under different PF with their respective training strategy:
GAQ was proposed by~\citet{jin2021graph}, on capturing inter-cell relation for antenna tilt optimization with scalability towards unseen size of topology. WCGCN was proposed by ~\citet{hu2019strategies}, for capturing relation between transceiving pairs in D2D communication, for link-level SINR estimation. GINE can be seen as more expressive WCGCN, and is proven to be a suitable backbone for SSL by \citet{hu2019strategies}. MLP is a baseline model that does not incorporate graph structure.
In addition, as pointed out by~\citet{chen2018stochastic}, QoS KPIs like SINR strongly depend on each cell's spatial configuration. We performed an ablation study towards introducing the augmented geometric features $O$ as additional node-feature/pretext-task in different PF beyond those included in $X$ (i.e., antenna azimuth, sector ID, etc). $O$ includes three augmented parameters denoting cell's spatial configuration, as illustrated in Fig.~\ref{fig:geom}~\footnote{We augment $O$ globally for all $e_{ij} \in E$, regardless if $v_i$ and $v_j$ have applied same or different (will not interfere each other in this case) carrier frequency.}: 
\begin{inparaenum}[\itshape (i) \upshape] 
    \item Interference Area (IA): denotes the searching area of possible interference.
    \item Interference Distance (ID): denotes an averaged distance from the interference source in the searching area.
    \item Interference Centric (IC): denotes an averaged location of measurement in the searching area.
\end{inparaenum}
$O$ is augmented based on the horizontal beam-width of the serving antenna, with a radius of 10 $km$~\footnote{Note: One could use more precise value for estimated coverage radius from crowd-sourcing. Here we used 10 $km$ as an upper bound of coverage in cell configuration design. Thus, we guarantee the generalization ability of the solution regardless of the dataset/area.}. 

Besides the performance evaluation, we proposed \textbf{a novel SSL PT framework for \ref{form:real} to tackle the problem with realistic data constrain}, as illustrated in Fig.~\ref{fig: proposed}. Instead of using $O$ as input feature, we considered a two-phased training framework: \begin{inparaenum} [\itshape \textbf{Phase} \upshape] \item PT: We consider IA or ID as pretext-task for PT. We first pre-train the selected backbone GNN against the designated pre-text task on edge-level. This selection of pretext-tasks is motivated by a series of work in road test measurement of signal propagation (e.g.~\citet{cho2022simultaneous} points out distance, coordinate and searching areas are key factor to calibrate simulation result to real-world measurement, while IC does not perform well as pretext-task in our experience) \item FT: We freeze the weight of backbone GNN, and apply newly initialized readout layers on downstream task. Such fine-tuning (FT) formulation is inspired by graph lifelong learning~\cite{galke2021lifelong}: GNN may able to generalize by taking over the knowledge gained in related previous tasks, even few ground truth is provided in downstream tasks. \end{inparaenum} In \textbf{Phase} PT, given the ground truth of select pretext-task being $O^z$, we introduce the loss function in Equ.~\ref{equ:ptloss}:
\begin{equation}
\label{equ:ptloss}
    \mathcal{L}_{SSL}(H,O^z) = \frac{1}{|E|}\sum_{(v_i,v_j \in E)} || \cos{(h_i,h_j)} - O^z_{i,j} ||^2
\end{equation}
where $|E|$ denotes the number of edges in the training data. $h_i,h_j \in H$ denotes the output node embedding after readout layers , for node $v_i,v_j$, respectively. In \textbf{Phase} FT, we adopt mean squared error (MSE) as loss function $\mathcal{L}_{MSE}(H,Z)$ for downstream task ground truth $Z$. We denote $\alpha\%$ as the percentage required for labeled data in \textbf{Phase} FT. To summarize, we present the pseudo-code in Algorithm~\ref{alg:SSL}. Note $[x||y]$
represents concatenating model weight dict $x$ and $y$.
\begin{algorithm}[ht]
\SetAlgoLined
\textbf{Parameter}: \\
\begin{myenumerate}
    \item Epoch length: $N_{PT}$,$N_{FT} \in \mathbb{N}$; percentage of ground truth needed: $\alpha\%$; training hyper-parameter for \textbf{Phase} PT and FT: $\mathcal{HP}_{PT}$,$\mathcal{HP}_{FT}$.
\end{myenumerate}
\textbf{Initialization}: \\
\begin{myenumerate}
    \item Backbone model $\mathsmaller{GNN}$, weight $w_{GNN}$;
    PT readout layers $\mathsmaller{RO}_{PT}$, weight $w_{PT}$; 
    and downstream readout layers $\mathsmaller{RO}_{FT}$, weight $w_{FT}$.\\
    \item  Prepare graph representation $G(V,E,X,Y,\_,\_)$, 
    pretext-task ground truth $O^z$ for all $v \in V$. 
    Uniformly sample $v_z \in V_z, V_z \subset V$ by $\alpha\%$. 
    Prepare downstream task ground truth $z \in Z$. 
    Optimizer $\nabla$.\\
\end{myenumerate}
\For{$e_{PT} = 1,\dots,N_{PT}$}{
\begin{myenumerate}
    \item $H = \mathsmaller{RO}_{PT}(\mathsmaller{GNN}(G(V,E,X,Y,\_,\_)))$
    \item $\nabla(\mathcal{L}_{SSL}(H,O^z), [w_{GNN}||w_{PT}],\mathcal{HP}_{PT})$
\end{myenumerate}
}
\begin{myenumerate}
    \item Frozen $w_{GNN}$.
\end{myenumerate}
\For{$e_{FT} = 1 \dots,N_{FT}$}{
\begin{myenumerate}
    \item $H = \mathsmaller{RO}_{FT}(\mathsmaller{GNN}(G(V,E,X,Y,\_,\_)))$
    \item $H_z \leftarrow$ Select $V_z$'s embeddings from $H$
    \item $\nabla(\mathcal{L}_{MSE}(H_z,Z), [w_{FT}],\mathcal{HP}_{FT} )$
\end{myenumerate}
}
\caption{Proposed GNN SSL Framework on FSL}
\label{alg:SSL}
\end{algorithm}
\section{Experiment}
\label{sec:exp}
We have trained our model against the real network configuration in two geographical locations, i.e., Copenhagen (CPH) and Aalborg-Aarhus (A+A).
The network measurement and KPIs are obtained from simulation in ~\texttt{Info-Vista Planet}~\footref{note1}.
\ref{form:full}: We first benchmark the state-of-the-art performance of GNNs benchmark in the Transductive setting, with an ablation study against $O$ as an additional node feature. The result in $MSE(\%)$ is presented in Table~\ref{table:pf1trans}. For the generalization setting, we denote $X\rightarrow~Y$ as training on dataset $X$ and test the model performance on dataset $Y$ to verify the model's generalization ability across different scenarios. Similarly, the generalization performance is shown in Table~\ref{table:pf1ind}.\\
\begin{table}
\caption{Model transductive Performance in MSE\,($\%$)}
 \centering 
 \label{table:pf1trans}
 \setlength{\tabcolsep}{3pt}
 \begin{threeparttable}
    \begin{tabular}{c|c|cc|cc}
    \toprule
    Model & KPI & CPH & CPH(w.$O$) & A+A & A+A(w.$O$) \\
    \midrule
    GAT   & SINR & 5.7 ± 0.5\% & 5.6 ± 0.5\% & 5.6 ± 0.2\% & 5.6 ± 0.1\%\\
    WCGCN & SINR & \textbf{3.2 ± 0.7\%} & \textbf{3.1 ± 0.3\%} & \textbf{1.1 ± 0.1\%} & \textbf{1.1 ± 0.2\%}\\
    GINE  & SINR& 3.4 ± 0.3\% & 3.5 ± 0.4\% & 2.0 ± 0.1\% & 1.8 ± 0.2\%\\
    MLP & SINR & 5.5 ± 0.1\% & 5.5 ± 0.1\% & 5.5 ± 0.1\% & 5.5 ± 0.1\%\\
    \midrule
    GAT & CQI& 0.31 ± 0.1\% & 0.30 ± 0.1\% & 2.9 ± 0.1\% & 2.9 ± 0.1\%\\
    WCGCN & CQI& \textbf{0.24 ± 0.2\%} & \textbf{0.24 ± 0.1\%} & \textbf{1.3 ± 0.2\%} & \textbf{1.2 ± 0.2\%}\\
    GINE & CQI& 0.28 ± 0.2\% & 0.26± 0.1\% & 1.7 ± 0.1\% & 1.6 ± 0.2\%\\
    MLP & CQI & 0.3 ± 0.1\% & 0.29 ± 0.1\% & 3.1 ± 0.1\% & 3.1 ± 0.1\%\\
    \bottomrule
    \end{tabular}
\end{threeparttable}
\end{table}
\begin{table}
\caption{Model inductive performance in MSE\,($\%$)}
\centering 
\label{table:pf1ind}
\setlength{\tabcolsep}{0.8pt}
    \begin{threeparttable}
        \begin{tabular}{c|c|cc|cc}
            \toprule
            \thead{Model} & \thead{KPI} & CPH$\rightarrow$A+A & CPH$\rightarrow$A+A(w.$O$) & A+A$\rightarrow$CPH & A+A$\rightarrow$CPH(w.$O$) \\
            \midrule
            GAT   & SINR & \textbf{9.2 ± 0.2\%} & \textbf{9.35 ± 0.3\%} & 11.3 ± 0.8\% & 11.2 ± 0.7\%\\
            WCGCN & SINR & 9.5 ± 0.3\% & 9.4 ± 1.7\% & 11.6 ± 0.3\% & \textbf{8.7 ± 0.3\%}\\
            GINE  & SINR & 9.4 ± 2.0\% & 11.5 ± 4.2\% & 12.3 ± 3.0\% & \textbf{8.7 ± 0.2\%}\\
            MLP   & SINR & 12.0 ± 0.2\% & 12.1 ± 0.9\% & \textbf{10.9 ± 0.7\%} & 10.9 ± 0.7\%\\
            \midrule
            GAT   & CQI & 3.7 ± 0.2\% & 3.5 ± 0.8\% & 0.39 ± 0.1\% & \textbf{0.45 ± 0.7\%}\\
            WCGCN & CQI & 4.3 ± 0.1\% & 4.05 ± 0.0\% & \textbf{0.34 ± 0.2\%} & 0.52 ± 0.3\%\\
            GINE  & CQI & 3.8 ± 0.7\% & 4.9 ± 2.9\% & 0.52 ± 0.4\% & 1.13 ± 0.9\%\\
            MLP   & CQI & \textbf{3.5 ± 0.4\%} & \textbf{3.5 ± 0.4\%} & 0.88 ± 0.4\% & 0.88 ± 0.4\%\\
            \bottomrule
        \end{tabular}
    \end{threeparttable}
\end{table}
Given presented result in Tables~\ref{table:pf1trans} and ~\ref{table:pf1ind}, one can conclude that, 
\begin{inparaenum}[\itshape (i) \upshape] For Table~\ref{table:pf1trans}: 
    \item WCGCN outperforms most benchmark models when given sufficient measurement on QoS KPI, while GAT has almost equivalent performance to MLP, which is aligned with the conclusion of~\citet{jin2021graph}. 
    \item Comparing the performance between `with' and `without' $O$ as additional edge attribute in Table~\ref{table:pf1trans}, we observe that the augmented geometric feature does not improve transductive performance with sufficient ground truth. 
    \item Comparing the respective cells between Table~\ref{table:pf1trans} and \ref{table:pf1ind}, we observe that the GNN model can not generalize its performance to new area/dataset, without degradation. This is aligned with the conclusion of \textbf{Appendix A} by \citet{yehudai2021local}, as the test dataset introduces unseen structural feature and graph size during training time. 
    \item By carefully studying the variance (e.g: GINE on CPH$\rightarrow$A+A) on the $MSE(\%)$ in Table~\ref{table:pf1ind}, we find that weight initialization in training time has a dominant impact on the inductive generalization performance toward another dataset. 
    \item Seeing Table~\ref{table:pf1ind}, introducing geometric feature (w.$O$ columns) does not benefit GNN generalization performance stably.
\end{inparaenum}\\
\ref{form:real}: Since \ref{form:full} conclude that given sufficient measurement on QoS KPI, the network geometric could be less important for GNNs.
We further studied the importance of $O$ in a realistic PF. In \ref{form:real}, we compared our proposed SSL framework with state-of-the-art training framework, as Fig.~\ref{fig: solutionrep}. We adjust provided percentage of ground truth $\alpha\%$ to verify the gain brought by \textbf{Phase} PT in SSL.
    
In Table~\ref{table:pf2trans}, we consider two different baselines: \begin{inparaenum} \item We consider no PT training scheme as a first baseline: if we applied a successful pretext-task, the GNN performance should outperform the same model without PT. We list the performance diff in column \textbf{\sc{Gain}}. \textcolor{ForestGreen}{Green text} denotes SSL result (i.e. $O\leftarrow$ID/IA) outperform the benchmark by a positive gain (i.e. $\Delta^{O}_{\text{CPH}}>0$), while \textcolor{red} {Red text} denotes a negative gain, indicating introducing pretext-task $O$ has down-graded performance.
    \item We consider MLP as a second baseline: Given MLP does not incorporate inter-cell relation, if any GNN model benefits from SSL more than MLP does with the same $\alpha\%$ (e.g: $\Delta^{O\text{, GINE}}_{\text{CPH}}$ $>$ $\Delta^{O\text{, MLP}}_{\text{CPH}}$), we consider the SSL framework is helping GNN backbone to incorporate inter-cell relation better, and color the block in green. 
\end{inparaenum}

\begin{table*}
\caption{Benchmark FSL Performance in MSE \,($\%$), given $\alpha~\%$ ground truth and pretext-task as $O$}
 \centering 
 \label{table:pf2trans}
 \setlength{\tabcolsep}{2pt}
 \begin{threeparttable}
    \begin{NiceTabular}{c|c|c|cccc|cccc}
    \toprule
    Model & KPI & $\alpha\%$ & 
    CPH & CPH(PT,$O\leftarrow$ID) & CPH(PT,$O\leftarrow$IA) & \sc{Gain} ($\Delta^{O}_{~\text{CPH}}$) &
    A+A & A+A(PT,$O\leftarrow$ID) & A+A(PT,$O\leftarrow$IA) & \sc{Gain} ($\Delta^{O}_{~\text{A+A}}$) \\
    \midrule
    \Block{3-1}{GAT} 
    & SINR & 2.5\% & 8.9 ± 0.2\% & 9.0 ± 0.1\% & 9.0 ± 0.1\% & \textcolor{red}{-0.1\%/-0.1\%}
    & 9.1 ± 0.1\% & 8.9 ± 0.1\% & 9.1 ± 0.0\% & \textcolor{ForestGreen}{+0.2\%}/0\\
    & SINR & 5\% & 8.9 ± 0.0\% & 8.9 ± 0.0\% & 8.9 ± 0.0\% & 0/0 &
    9.0 ± 0.0\% & 8.9 ± 0.0\% & 8.9 ± 0.1\% & \textcolor{ForestGreen}{+0.1\%/+0.1\%}\\
    & SINR & 10\% & 6.2 ± 0.0\% & \cellcolor{green!25}6.0 ± 0.0\% & 6.2 ± 0.0\% & \textcolor{ForestGreen}{+0.2\%}/0 
    & 8.8 ± 0.0\% & 8.8 ± 0.1\% & 8.9 ± 0.0\% & 0/\textcolor{red}{-0.1\%}\\
    \midrule
    \Block{3-1}{WCGCN} 
    & SINR & 2.5\% & 6.6 ± 0.2\% & 6.5 ± 0.2\% & \cellcolor{green!25}6.0 ± 0.2\% & \textcolor{ForestGreen}{+0.1\%/+0.6\%}
    & 6.8 ± 0.1\% & 7.1 ± 1.0\% & 6.9 ± 0.3\% & \textcolor{red}{-0.3\%/-0.1\%}\\
    & SINR & 5\% & 5.9 ± 0.2\% & 6.4 ± 0.3\% & \cellcolor{green!25}5.4 ± 0.1\% & \textcolor{red}{-0.5\%}/\textcolor{ForestGreen}{+0.5\%}
    & 5.9 ± 0.1\% & 6.7 ± 1.3\% & 6.3 ± 0.4\% & \textcolor{red}{-0.8\%/-0.4\%}\\
    & SINR & 10\% & 5.4 ± 0.5\% & 5.3 ± 0.1\% & \cellcolor{green!25}5.0 ± 0.1\% & \textcolor{ForestGreen}{+0.1\%/+0.4\%}
    & 5.3 ± 0.1\% & 5.7 ± 0.1\% & \cellcolor{green!25}5.0 ± 0.1\% & \textcolor{red}{-0.4\%}/\textcolor{ForestGreen}{+0.3\%}\\
    \midrule
    \Block{3-1}{GINE}
    & SINR & 2.5\% & 7.6 ± 0.8\% & \cellcolor{green!25}6.0 ± 0.2\% & \cellcolor{green!25}6.2 ± 0.2\% & \textcolor{ForestGreen}{+1.6\%/+1.4\%}
    & 7.8 ± 0.4\% & \cellcolor{green!25}5.9 ± 0.1\% & \cellcolor{green!25}6.6 ± 0.2\% & \textcolor{ForestGreen}{+1.9\%/+1.2\%}\\
    & SINR & 5\% & 6.9 ± 0.4\% & \cellcolor{green!25}5.9 ± 0.1\% & \cellcolor{green!25}5.4 ± 0.1\% & \textcolor{ForestGreen}{+1\%/+1.5\%} & 6.3 ± 0.2\% & \cellcolor{green!25}5.9 ± 0.1\% & \cellcolor{green!25}5.7 ± 0.2\% & \textcolor{ForestGreen}{+0.4\%/+0.6\%}\\
    & SINR & 10\% & 5.5 ± 0.2\% & \cellcolor{green!25}5.0 ± 0.1\% & \cellcolor{green!25}5.2 ± 0.1\% & \textcolor{ForestGreen}{+0.5\%/+0.3\%}
    & 6.1 ± 0.3\% & \cellcolor{green!25}5.6 ± 0.2\% & \cellcolor{green!25}5.5 ± 0.0\% & \textcolor{ForestGreen}{+0.5\%/+0.6\%}\\
    \midrule
    \Block{3-1}{MLP}
    & SINR & 2.5\% & 9.1 ± 0.1\% & 9.0 ± 0.1\% & 9.1 ± 0.1\% & \textcolor{ForestGreen}{+0.1\%}/0
    & 9.1 ± 0.2\% & 8.9 ± 0.0\% & 8.9 ± 0.0\% & \textcolor{ForestGreen}{+0.2\%/+0.2\%}\\
    & SINR & 5\% & 8.9 ± 0.0\% & 8.9 ± 0.0\% & 8.9 ± 0.0\% & 0/0
    & 9.0 ± 0.0\% & 8.9 ± 0.0\% & 8.9 ± 0.0\% & \textcolor{ForestGreen}{+0.1\%/+0.1\%}\\
    & SINR & 10\% & 6.7 ± 0.0\% & 6.7 ± 0.1\% & 6.7 ± 0.1\% & 0/0 
    & 9.0 ± 0.2\% & 8.9 ± 0.1\% & 8.9 ± 0.1\% & \textcolor{ForestGreen}{+0.1\%/+0.1\%}\\
    \midrule
    \Block{3-1}{GAT}
    & CQI & 2.5\% & 0.50 ± 0.0\% & 0.52 ± 0.0\% & 0.50 ± 0.0\% & \textcolor{ForestGreen}{+0.02\%}/0  
    & 3.3 ± 0.0\% & 3.3 ± 0.0\% & 3.4 ± 0.0\% & 0/\textcolor{red}{-0.1\%}\\
    & CQI & 5\% & 0.40 ± 0.0\% & 0.40 ± 0.0\% & 0.40 ± 0.0\% & 0/0
    & 3.3 ± 0.0\% & 3.3 ± 0.0\% & 3.3 ± 0.0\% & 0/0\\
    & CQI & 10\% & 0.30 ± 0.0\% & 0.30 ± 0.0\% & 0.30 ± 0.0\% & 0/0
    & 3.3 ± 0.1\% & 3.3 ± 0.1\% & 3.3 ± 0.0\% & 0/0\\
    \midrule
    \Block{3-1}{WCGCN} 
    & CQI & 2.5\% & 0.50 ± 0.1\% &  0.46 ± 0.0\% & 0.40 ± 0.0\% & \textcolor{ForestGreen}{+0.04\%/+0.1\%} 
    & 3.1 ± 0.1\% & 3.2 ± 0.1\% & 3.3 ± 0.0\% & \textcolor{red}{-0.1\%/-0.2\%}\\
    & CQI & 5\% & 0.40 ± 0.0\% & 0.28 ± 0.0\% & 0.31 ± 0.0\% & \textcolor{ForestGreen}{+0.12\%/+0.09\%}
    & 3.0 ± 0.0\% & 3.1 ± 0.1\% & 3.3 ± 0.2\% & \textcolor{red}{-0.1\%/-0.3\%}\\
    & CQI & 10\% & 0.40 ± 0.0\% & 0.25 ± 0.0\% & 0.25 ± 0.0\% & \textcolor{ForestGreen}{+0.15\%/+0.15\%}
    & 3.1 ± 0.0\% & 3.0 ± 0.1\% & 3.3 ± 0.2\% & \textcolor{ForestGreen}{+0.1\%}/\textcolor{red}{-0.2\%}\\
    \midrule
    \Block{3-1}{GINE}
    & CQI & 2.5\% & 3.1 ± 0.5\% & \cellcolor{green!25}0.33 ± 0.1\% & \cellcolor{green!25}0.30 ± 0.0\% & \textcolor{ForestGreen}{+2.8\%/+2.8\%} & 3.6 ± 0.1\% & \cellcolor{green!25}2.8 ± 0.0\% & \cellcolor{green!25}3.0 ± 0.1\% & \textcolor{ForestGreen}{+0.8\%/+0.6\%}\\
    & CQI & 5\% & 2.3 ± 0.3\% & \cellcolor{green!25}0.33 ± 0.0\% & \cellcolor{green!25}0.28 ± 0.0\% & \textcolor{ForestGreen}{+2.0\%/+2.0\%}
    & 3.4 ± 0.1\% & \cellcolor{green!25}2.9 ± 0.0\% & \cellcolor{green!25}2.8 ± 0.0\% & \textcolor{ForestGreen}{+0.5\%/+0.6\%}\\
    & CQI & 10\% & 1.2 ± 0.3\% & \cellcolor{green!25}0.24 ± 0.0\% & 0.22 ± 0.0\% & \textcolor{ForestGreen}{+0.96\%/+0.98\%} 
    & 3.2 ± 0.1\% & \cellcolor{green!25}2.7 ± 0.0\% & \cellcolor{green!25}2.7 ± 0.0\% & \textcolor{ForestGreen}{+0.5\%/+0.5\%} \\
    \midrule
    \Block{3-1}{MLP}
    & CQI & 2.5\% & 1.4 ± 0.1\% & 0.40 ± 0.0\% & 0.30 ± 0.0\% & \textcolor{ForestGreen}{+1.0\%/+1.1\%}
    & 3.7 ± 0.1\% & 3.3 ± 0.0\% & 3.4 ± 0.0\% & \textcolor{ForestGreen}{+0.4\%/+0.3\%} \\
    & CQI & 5\% & 1.4 ± 0.3\% & 0.40 ± 0.0\% & 0.40 ± 0.0\% & \textcolor{ForestGreen}{+1.0\%/+1.0\%}
    & 3.6 ± 0.1\% & 3.4 ± 0.0\% & 3.3 ± 0.0\% & \textcolor{ForestGreen}{+0.2\%/+0.3\%} \\ 
    & CQI & 10\% & 1.4 ± 0.3\% & 0.50 ± 0.0\% & 0.40 ± 0.0\% & \textcolor{ForestGreen}{+0.9\%/+1.0\%}
    & 3.6 ± 0.1\% & 3.4 ± 0.0\% & 3.3 ± 0.0\% & \textcolor{ForestGreen}{+0.2\%/+0.3\%} \\ 
    \bottomrule
    \end{NiceTabular}
\end{threeparttable}
\end{table*}
Given the presented result in Table~\ref{table:pf2trans}, one can conclude: \begin{inparaenum}[\itshape (i) \upshape] 
    \item GNN degrades heavily with insufficient ground truth compared with~\ref{form:full}. When compared with the second baseline (MLP), one can see degradation is not only due to difficulty in generalizing towards inter-cell relational features but also node attributes, as MLP also improves its SSL performance for CQI task.
    \item When compared with the first baseline (train without SSL), SSL PT benefits GNNs unevenly on both inter-cell relational feature and node attributes. \item For the QoS KPI estimation task, a more expressive GNN backbone benefits more from PT, than less expressive ones (e.g., GINE is considered to be expressive by~\citet{hu2019strategies}, while WCGCN is less expressive due to an additional \textit{maxpool} during message passing beyond GINE's design~\cite{shen2020graph}, expressiveness: GINE $>$ WCGCN $>$ GAT). This is aligned with the conclusion by~\citet{hu2019strategies} who observed similar behavior in chemistry \& biology datasets. However, their solution is limited to small, non-planar graphs while our work extends the conclusion for QoS KPI estimation, where the graph is near-planar.
    Furthermore, we cherry-pick the best-performing model in FSL by dataset and their estimating KPI to compare against the same model's performance in full training in table~\ref{table:pf2conclu}, where 'SOTA' denotes a full training using 80\% of ground truth (aligned with Fig.~\ref{fig: proposed}, \textit{State-of-the-art} training scheme). 
    This supports our augments that a proper SSL framework with only 2.5\% ground truth can compete with 'SOTA' training in many downstream tasks. 
\end{inparaenum} 
We consider that our pre-text task (IA/ID) helps our GNN models to learn network geometry knowledge, which enables them to outperform the state-of-the-art training framework. Finally, we include a comparative result in table~\ref{tab:runtimecomp} between GNN solution and the baseline simulator \textbf{\sc{InfoVista Planet}} to prove GNN is more time and computationally efficient. Seeing from the RAM and inference time, GNN is significantly superior to the baseline simulator. Since inference time for GNN is cell-independent, we exclude the running time for the simulator for preprocessing per-cell's configuration parameter as that can grow linearly with the number of cells for a fair comparison. (This step could take additional 2 hours for an A+A dataset.) One can also see from search space size: In dataset (A+A) with more rural components, each cell's QoS estimation is less dependent on its neighboring cell than the 'more urban' one (CPH). Both methods reflect that, while GNN more aggressively reduces the search space, this accelerates its inference, but degrades its performance in more urban areas. 
\begin{table}
\caption{Comparing Best Performing FSL ($\alpha = 2.5\%$) with SSL on Each Task, against Full Training (SOTA), in MSE($\%$) }
 \centering 
 \label{table:pf2conclu}
 \setlength{\tabcolsep}{2pt}
 \begin{threeparttable}
    \begin{NiceTabular}{c|c|c|c|cccc}
    \toprule
    Dataset & KPI & $\alpha\%$ & Best Model & $O$ & FSL MSE(\%) & SOTA MSE(\%) & Gap\\
    \midrule
    \multirow{2}{*}{CPH} & \multirow{2}{*}{SINR} & 2.5\% & WCGCN & IA & 6.0 ± 0.2\% & 4.0 ± 0.2\% & -2.0\%\\
                        &   & 2.5\% & GINE & ID & 6.0 ± 0.2\% & 4.3 ± 0.2\% & -1.7\%\\
    CPH & CQI & 2.5\% & GINE & IA & 0.30 ± 0.0\% & 0.8 ± 0.3\% & 0.5\%\\ 
    A+A & SINR & 2.5\% & GINE & ID & 5.9 ± 0.1\% & 5.6 ± 0.1\% & -0.3\%\\
    A+A & CQI & 2.5\% & GINE & ID & 2.8 ± 0.0\% & 2.8 ± 0.1\% & 0 \\
    \bottomrule
    \end{NiceTabular}
\end{threeparttable}
\end{table}
\begin{table}[ht]
\centering
\caption{Computational Resource Comparison}
\label{tab:runtimecomp}
 \setlength{\tabcolsep}{3pt}
\begin{threeparttable}
\begin{tabularx}{\linewidth}{p{0.27\linewidth}p{0.15\linewidth}p{0.15\linewidth}p{0.125\linewidth}p{0.125\linewidth}}
\toprule
Model & \multicolumn{2}{c}{GNN (GINE/WCGCN)} & \multicolumn{2}{c}{\textbf{\sc{InfoVista Planet}}} \\
\cmidrule(lr){2-3} \cmidrule(lr){4-5}
Dataset & CPH & A+A & CPH & A+A \\
\hline
RAM &  $\leq 15$ MiB & $\leq 15$ MiB & $1.2$ GiB & $1.65$ GiB\\
Inference Runtime & $\leq 0.2s$ & $\leq 0.2s$ & 13.18 min & 36.72 min\\ 
Searchspace per result & $8.456$ & $7.612$  & $54.9$ & $14.35$\\
\bottomrule
\end{tabularx}
\end{threeparttable}
\end{table}
\section{Conclusion}
\label{sec:conclusion}
The paper evaluates state-of-the-art coverage-estimation-targeted GNNs in real-world configurations and identifies inefficiencies in generalization when transferring learning outcomes from one scenario to another. In addition, we formulate a more realistic few-shot learning problem that considers data constraints and propose a novel pretext-task for PT to facilitate FSL performance with minimal performance degradation. Our training framework confirms the benefits of SSL with fewer ground truth samples. Future work can focus on improving the model's generalization ability and exploring additional zero-shot estimation schemes.
\section*{Acknowledgement}
This work was partially supported by the Wallenberg AI, Autonomous Systems and Software Program (WASP) funded by the Knut and Alice Wallenberg Foundation.


\printbibliography

\medskip


\end{document}